\documentstyle[preprint,tighten,prb,aps]{revtex}

\draft
\newcommand{\ab}{\textit{ab initio }}

\newcommand{\Ca}{CaMnO$_3$ }
\newcommand{\Cap}{CaMnO$_3$}
\newcommand{\La}{LaMnO$_3$ }
\newcommand{\Lap}{LaMnO$_3$}

\newcommand{\CaLap}{Ca$_{1-x}$La$_x$MnO$_3$}
\newcommand{\LaCu}{La$_{2}$CuO$_4$}
\newcommand{\pnma}{\textit{Pnma }}

\newcommand{\eg}{e$_g$ }
\newcommand{\tg}{t$_{2g}$ }

\newcommand{\dxx}{d$_{x^{2}-y^{2}}$ }

\newcommand{\dzz}{d$_{3z^{2}-r^{2}}$ }
\newcommand{\dx}{d$_{3x^{2}-r^{2}}$ }
\newcommand{\dy}{d$_{3y^{2}-r^{2}}$ }
\newcommand{\oion}{O$^{2-}$ }
\newcommand{\oionp}{O$^{2-}$}
\newcommand{\mmm}{Mn$^{3+}$ }
\newcommand{\mmmm}{Mn$^{4+}$ }

\newcommand{\mcca}{Mn$_2$O$_{11}^{14-}$ }
\newcommand{\mcla}{Mn$_2$O$_{11}^{16-}$ }
\newcommand{\mclap}{Mn$_2$O$_{11}^{16-}$}

\newcommand{\jpar}{J$_{\|}$ }
\newcommand{\jper}{J$_{\bot}$ }
\newcommand{\jparp}{J$_{\|}$}
\newcommand{\jperp}{J$_{\bot}$}
\newcommand{\jone}{J$_{1}$ }
\newcommand{\jtwo}{J$_{2}$ }

\newcommand{\jonep}{J$_{1}$}
\newcommand{\jtwop}{J$_{2}$}

\newcommand{\etal}{\textit{et al.} }
\newcommand{\ie}{\textit{i.e. }}
\newcommand{\etc}{\textit{etc. }}
\newcommand{\etcp}{\textit{etc.}}
\newcommand{\cf}{\textit{c.f. }}

\begin{document}

\bibliographystyle{prsty}

\title{Exchange coupling in \Ca and \Lap: configuration interaction and the coupling mechanism}
\author{M. Nicastro and C.H. Patterson}
\address{Department of Physics and Centre for Scientific Computation,\\
University of Dublin, Trinity College, Dublin 2, Ireland.}
\date{\today}
\maketitle

\begin{abstract}

The equilibrium structure and exchange constants of \Ca and \La have been investigated using total energy unrestricted
Hartree-Fock (UHF) and localised orbital configuration interaction (CI) calculations on the bulk compounds and \mcca and \mcla clusters.
The predicted structure and exchange constants for \Ca are in reasonable agreement with estimates based on its N\'eel temperature.
A series of calculations on \La in the cubic perovskite structure shows that a Hamiltonian with independent orbital ordering 
and exchange terms accounts for the total energies of cubic \La with various spin and orbital orderings. Computed exchange constants
depend on orbital ordering. Exchange contributions to the total energy vary between -20 and 20 meV per Mn ion, differences in
orbital ordering energy vary between 3 and 100 meV and a Jahn-Teller distortion results in an energy reduction of around 300 meV. 
The lattice constant of the lowest energy cubic perovskite structure (3.953 \AA) is in good agreement with the lattice constant of
the high temperature 'cubic' phase of \La (3.947 \AA). The total energy of \pnma \La was minimised by varying lattice parameters and seven
internal coordinates and a structure 194 meV per Mn ion below that of a structure determined by neutron diffraction was found. This optimised
structure is nearly isoenergetic with a cubic perovskite structure with a 5 percent Jahn-Teller distortion.
UHF calculations tend to underestimate exchange constants in \Lap,  but have the correct sign when compared with values obtained
by neutron scattering; exchange constants obtained from CI calculations are in good agreement with neutron
scattering data provided the Madelung potential of the cluster is appropriate.
Cluster CI calculations reveal a strong dependence of exchange constants on Mn d \eg orbital populations in both compounds.
CI wave functions are analysed in order to determine which exchange processes are important in exchange coupling in \Ca and \Lap.

\end{abstract}

\pacs{71.27.+a 75.10.-b 75.50.-y}

%\twocolumn
\section{Introduction}

\Ca and \La are endpoint compounds in the series \CaLap, which has been thoroughly studied experimentally and theoretically \cite{Coey99}.
They have relatively simple atomic and magnetic structures, their magnetic excitations are well described by a 
spin wave Hamiltonian \cite{Endoh96,Moussa96} and exchange constants, J, are well established by neutron scattering \cite{Endoh96,Moussa96} 
and from the N\'eel temperature \cite{Koehler55,Rushbrooke74}. 
Exchange coupling in manganites has been extensively studied using model Hamiltonian \cite{Millis97,Feiner99,Satpathy00}
and \ab calculations \cite{Terakura95,Pickett96,Terakura96,Terakura96a,Terakura99,Patterson00,Su00}. 
This paper presents results of bulk \ab Unrestricted Hartree-Fock (UHF) 
 and cluster Configuration Interaction (CI) calculations of exchange constants for both compounds. 

Exchange constants obtained from CI calculations are in excellent agreement with experiment
and the localised orbital CI wavefunctions are analysed to determine which quantum fluctuations are most important
in exchange coupling.
Model Hamiltonian calculations have attributed the exchange coupling energy to \oion to Mn$^{3+/4+}$ superexchange 
\cite{Millis97}, Mn~d$^{4+}$~d$^{4+}$/Mn~d$^{5+}$~d$^{3+}$ \tg superexchange \cite{Feiner99}, or both \cite{Satpathy00}.
Results of calculations presented below show that both exchange mechanisms operate and that O superexchange is
the more important of the two. This was also found to be the case in the model Hamiltonian calculations of 
Meskine \etal \cite{Satpathy00}.

CI cluster calculations provide detailed information on exchange couplings between neighbouring Mn ions, however a bulk electronic structure
technique is required to study orbital ordering in \Lap. Total energies of \La with A and G-type
anti-ferromagnetic (A-AF and G-AF) and ferromagnetic (FM) spin orderings have been computed in several isovolume structures in order to 
establish whether or not orbital ordering and spin ordering terms in the Hamiltonian for \La are independent.
Obviously exchange constants will depend on orbital ordering, as the latter determines which empty orbitals
are available to participate in exchange coupling. However, it is not known whether the \eg electron density in \La for a particular orbital
ordering depends on spin ordering. It is shown below that a common orbital ordering energy for any of several orbital orderings
can be identified
and that this energy is independent of spin ordering, to a high degree. Spin and orbital ordering terms in the Hamiltonian are therefore
independent, although orbital ordering determines the exchange constants.

At low temperatures, \Ca exists in a cubic perovskite structure (lattice constant 3.73~\AA) with G-AF magnetic ordering \cite{Koehler55}
and a N\'eel temperature of 130~K. Using the Rushbrooke-Wood formula \cite{Rushbrooke74}, this N\'eel temperature implies an 
exchange constant J = 6.6~meV. Note that throughout this work the spin Hamiltonian is of the form due to Domb and Sykes
\cite{Domb74}

\begin{equation}\label{eqn:eqn1}
H = \sum_{<ij>}{}  J_{ij} \frac{\hat{S}_{i}.\hat{S}_{j}}{S^{2}}
\end{equation}

${\hat{S}_{i}}$ is a spin operator, S is the magnitude of the total spin for an ion and J$_{ij}$ is the exchange constant
for a pair of ions. This form is adopted for the Hamiltonian as it is the same as that adopted in modelling spin wave dispersion in
neutron scattering studies \cite{Endoh96,Moussa96}, except for a small Dzyaloshinsky-Moriya term.

At low temperatures
the space group of \La is \pnma \cite{Koehler55}. The ground state magnetic structure is A-AF
and the unit cell contains four formula units consisting of rotated and distorted octahedra.  
There is one more d electron per Mn ion (${\it{c.f.}}$ \Cap), which occupies an \eg orbital and induces a Jahn-Teller distortion 
in each MnO$_{6}$ octahedron, resulting in three distinct Mn-O bond lengths of 1.91, 1.97 and 2.18~\AA \cite{Moussa98}. 
The occupied \eg orbital is a linear combination of \dxx and \dzz orbitals.
The largest orbital component lies along the most elongated Mn-O bond. 
The \pnma structure is shown in 
Fig. \ref{fig:fig1}. $a$, $b$ and $c$ axes referred to below are indicated on this diagram.
The results of a number of neutron and x-ray scattering studies of the structure of \La 
\cite{Elemans71,Tofield74,Roosmalen91,Norby95,Mitchell96,Huang97} over a range of temperatures are summarised in Ref.{\cite{Moussa98}}. 
The \pnma structure can be viewed as containing planes of \mmm ions, each joined to its in-plane neighbours by pairs of short (1.91~\AA) and
long (2.18~\AA) Mn-O bonds. Each \mmm ion in a particular plane is coupled to \mmm ions in planes immediately above and below by
two Mn-O bonds (1.97~\AA). The symmetry of the \pnma structure is such that there is one in-plane (nearest neighbour)
exchange constant (\jpar)
and one out-of-plane constant (\jper).
Both \jpar and \jper have been determined from two independent neutron scattering studies to be
-6.7 and 4.8 meV, respectively, \cite{Endoh96,Moussa96}. Thus there is FM coupling within planes and AF coupling between planes.

\section{Exchange Coupling Mechanism}

The first comprehensive attempt to explain atomic and magnetic structures in doped and undoped manganites was made by Goodenough 
\cite{Goodenough55} in 1955. 
He assumed three classes of exchange interaction between neighbouring Mn ions in undoped
\Ca and \La lattices. A specific ordering of $\textit{empty }$\eg orbitals and relative orientations of pairs of empty \eg orbitals
correponding to Goodenough's classification are illustrated in Fig.~\ref{fig:fig2}. 
When empty \eg orbitals are available on a pair of neighbouring Mn ions and are oriented towards one another
(Fig.~\ref{fig:fig2}b)
then AF coupling of the Mn ion spins is energetically favoured. This is because
electrons from the central \oion ion of either spin are postulated to delocalise onto both Mn ions simultaneously, owing to the favourable
exchange interaction (Hund's rule) between the delocalised electron and the Mn ion spin. 
However, if the Mn ion spins are FM aligned, only the electron from the central \oion ion with the same spin orientation as the Mn ions can
delocalise onto either Mn ion, resulting in a higher energy for that state.
Thus the empty orbital arrangement shown in Fig.~\ref{fig:fig2}b results in an AF coupling of Mn spins. This is a type I exchange interaction 
according to Goodenough \cite{Goodenough55}.
When one empty \eg orbital is suitably oriented for \oion ion electron delocalisation (Fig.~\ref{fig:fig2}c), 
FM coupling of the Mn ion spins is favoured.
This is a type II interaction.
Finally, when no empty hybrids are available (Fig.~\ref{fig:fig2}d), no delocalisation occurs. This is a type III interaction. 
This model has been used to
explain the relative energies of A-AF, G-AF and FM magnetic states of \Ca and \La with a cubic perovskite structure
\cite{Patterson00}. In that work it was found that the relative energies of these magnetic structures could be explained by 
counting the numbers of each type of interaction in each magnetic state and calculating the relative energy of each type of interaction.
For both \Ca and \La it was found that the type I AF interaction was more energetically favourable than the type II interaction by
10~meV \cite{Patterson00}. 
The simplified description of exchange interactions just given assumes that an empty \eg orbital is either available or not. However,
empty \eg orbitals in \La are not purely \dxx or \dzz in character \cite{Millis97}. The mixed character of the empty \eg orbital
permits some exchange coupling even when the empty \eg orbital is not optimally oriented.

This type of reasoning was used by Millis \cite{Millis97} in a calculation of exchange coupling energies in \Ca and \Lap. In that work
configurations allowed by the Pauli principle in which one or two electrons hop from the central \oion ion to one or both neighbouring
Mn ions are considered. Configurations which differ by a single electron hop have a single hopping matrix element, t. Diagonal elements
of the Hamiltonian are parametrised using the energy required to excite one electron or a pair of electrons from an \oion ion to a Mn ion.
In \Ca the configuration which is assumed to lead to stabilisation of the AF state over the FM state is 
one in which a pair of electrons on the \oion ion
is excited onto separate \mmmm ions. If this were indeed the origin of exchange coupling in \Ca then one would expect this configuration
to appear in an \ab ground state CI wave function, but this is not the case. 
However, the main idea of this model, that more low energy configurations
are available to singlet states than high spin multiplicity states, is in accord with results presented here.

CI cluster calculations of exchange constants in
La$_{2}$CuO$_{4}$ \cite{Martin93,Martin97} and
KNiF$_{3}$ \cite{Illas97}, 
which used delocalised molecular orbitals have been reported quite recently.
CI cluster calculations described below were carried out in a localised orbital basis. The localised orbital basis provides a means of 
identifying the exchange coupling mechanism in terms of fluctuations of electrons between localised orbitals.
These calculations were performed on \mcca and \mcla clusters representing fragments of \Ca and \Lap. 
Details of the calculations, including the method used to generate the localised orbitals, details of 
a spherical array of point charges surrounding the clusters, \etc are given in Appendix\ \ref{app:details}.
The wavefunctions for the clusters contain orbitals which are partitioned into a (doubly-occupied) core orbital space, an active
space containing the 2p orbitals of the \oion ion situated between the two Mn ions in the cluster as well as singly-occupied Mn d orbitals
and an external space containing unoccupied Mn d orbitals. The core orbital space contains 'core' electrons as well as valence electrons
not in the active orbital space. The ions in the clusters treated quantum mechanically consisted of two corner sharing MnO$_6$ octahedra.
The localised orbitals
in the active space for the \mcca and \mcla clusters are shown in Figs. \ref{fig:fig3} and \ref{fig:fig4}, respectively.
The main electronic configuration for the \mcca cluster representing \Ca is one in 
which each Mn ion with a formal 4+ charge contains three \tg electrons and each oxygen ion exists in a closed shell \oion configuration.
The actual charge on the Mn ions is significantly reduced as there is a covalent component to the Mn-O bonds, as can clearly be seen
in the contour plot of the localised orbital with mainly O 2pz character in the top panel of Fig. \ref{fig:fig3}. The actual Mn ion charge
in \Ca is +2.13, according to a Mulliken population analysis of the UHF wave functions obtained for \Cap. The formal charge on Mn ions
in \La is 3+, however a Mulliken population analysis of UHF wave functions for \La yields a charge of +2.24
The O ion charges in the two compounds are -1.33 (\Cap) and -1.75 and -1.82 (\Lap).
Exchange constants were calculated by finding the energy difference between the spin singlet and spin septet(nonet) states
of the \mcca and \mcla clusters. 

Wave functions were constructed from the localised orbitals shown in Figs. \ref{fig:fig3} and \ref{fig:fig4} and doubly occupied
core orbitals. A septet state for the \mcca cluster was constructed from 
six singly-occupied \tg orbitals and doubly occupied O 2p orbitals localised on the central O ion in the cluster. The form of this
wave function is 

\begin{equation}\label{eqn:eqn2}
\psi^{septet} = A\Big(\big\{core\big\}(\phi_{xy,l}\phi_{xz,l}\phi_{yz,l}\phi_{xy,r}\phi_{xz,r}\phi_{yz,r})\\
                  (\alpha\alpha\alpha\alpha\alpha\alpha)\Big)
\end{equation}

$A$ is the anti-symmetrising operator and the subscripts {\textit{l}} or {\textit{r}} on \tg orbitals in Eq.\ \ref{eqn:eqn2}
indicate that they are centred on the left or right Mn ion, respectively. $\big\{core\big\}$ is a 
product of doubly occupied orbitals in the core orbital space which includes the 2p orbitals on the central \oionp. 
This is the restricted open shell Hartree-Fock (ROHF) wave function for
the septet state, constructed using localised molecular orbitals. 
Self-consistent field (SCF) ROHF wave functions can be computed using a number of standard electronic structure
packages such as the GAMESS package \cite{GAMESS} used in this work.

The singlet state is constructed from the same set of singly-occupied orbitals with a spin-coupling of the form

\begin{equation}\label{eqn:eqn3}
                  \frac{1}{\sqrt{2}}(\alpha\alpha\alpha\beta\beta\beta-\beta\beta\beta\alpha\alpha\alpha)
\end{equation}

This is one of five spin eigenfunctions \cite{fermion} (SEF) for six electrons coupled into a singlet state. 
Provided that the spatial orbitals multiplying this SEF are ordered
such that orbitals localised on each Mn ion are grouped together, we expect this SEF to dominate the CI wave function, since
Hund's rule requires spins on each ion to be coupled with the same spin. This is indeed found to be the case in the actual CI wave 
function for the singlet state of the \mcca cluster representing \Cap. 
The wave function for the singlet state is therefore

\begin{equation}\label{eqn:eqn4}
\psi^{singlet} = \frac{1}{\sqrt{2}}A\Big(\big\{core\big\}(\phi_{xy,l}\phi_{xz,l}\phi_{yz,l}\phi_{xy,r}\phi_{xz,r}\phi_{yz,r})\\
                  (\alpha\alpha\alpha\beta\beta\beta-\beta\beta\beta\alpha\alpha\alpha)\Big)
\end{equation}

Using conventional rules for evaluating determinantal
energies \cite{Szabo}, the energy difference
between the two states is K$_{xz,l~xz,r}$ + K$_{yz,l~yz,r}$, with the singlet state lying {\textit{above}} the septet state
(assuming that other inter-site exchange integrals are zero because of negligible spatial overlap). 
When the ground state energies of the singlet and septet states of the \mcca cluster with wave functions in Eq. \ref{eqn:eqn2} and 
\ref{eqn:eqn4} were evaluated, the singlet state was 3.6~meV above the septet state. This implies a value of 1.8~meV for the exchange
integrals just mentioned. Note that we use the notation K$_{ij}$ for exchange integrals between specific molecular orbitals while we use
the notation J for the (effective) exchange coupling energy of two spins on different Mn ions.
The singlet and septet states of this configuration are analogous to the
Heitler-London valence bond wave function for the singlet and triplet states of the He atom in a 1s2s configuration. 
In that case the triplet state is lower than the singlet state by
K$_{1s2s}$.

In general, CI wave functions with N electrons in the active orbital space consist of linear combinations of 
spin-adapted functions (SAF's)

\begin{equation}\label{eqn:eqn5}
\psi_{CI} = \sum_{i} c_{i}~\psi^{SAF}_{i}
\end{equation}

\begin{equation}\label{eqn:eqn6}
\psi^{SAF}_{i} = A\Big(\big\{core\big\}\phi_{j}\phi_{k}...\phi_{s}\phi_{t} \Theta_{a}\Big)
\end{equation}
 
where a SAF is a product of spatial orbitals 
and a SEF, $\Theta_{a}$, for the particular spin state in question. 
The septet and singlet
SAF's in Eq. \ref{eqn:eqn2} and \ref{eqn:eqn4} are the dominant terms in a more general CI
expansion of the septet and singlet wave function of the \mcca cluster. All SAF's which are obtainable by exciting
one or two electrons from the dominant SAF's to empty orbitals in the active space are included in the expansion. As stated above,
the 13 orbitals in the active space in the calculations described here are comprised of 10 orbitals of mainly 
Mn 3d character and 3 of mainly O 2p character
localised on the O ion between the two Mn ions. These excited electron SAF's enter the wave function with a maximum weight of order 10$^{-2}$ 
and a corresponding occupancy of order 10$^{-4}$ and it is these which lower the energy of the singlet state below the septet state
when the spins are AF coupled. The main excited SAF's in the singlet and septet wave functions are those in which: $\textit{one}$ electron is 
excited from an O 2p orbital to the Mn \eg orbital aligned with the Mn-O axis (O to \eg (1e)); a $\textit{pair}$ of electrons are excited from
one O 2p orbital to the $\textit{same}$ Mn \eg orbital (O to \eg (2e)) and an excitation in which a 
\tg electron is transferred fron one Mn ion to the other
(\tg exchange).  Obviously the latter exchange process is only allowed in the singlet state as it violates the Pauli exclusion principle
in the septet state when the \tg shells are half filled, as in \Cap. Excitations in which a pair of electrons are excited from the O ion
to separate Mn ions are found to have negligible weight for both spin states. 

\section{Results}
\label{sec:RES}

\subsection{\Cap: Bulk UHF Calculations}

UHF total energy calculations were performed using the \textsc{crystal98} code \cite{crystal98} for FM, A-, C-, and G-AF spin orderings.
The energy of the cubic FM structure with the experimental lattice constant of 3.73~\AA~was adopted as the reference energy (0 meV);
calculations were also performed for each of the spin orderings with a lattice constant of 3.75~\AA.
Total energies and magnetic moments from 
these calculations are given in Table~\ref{tab:tab1}.
When these total energy differences are fitted to the Hamiltonian in Eq. \ref{eqn:eqn1} with nearest (\jonep) and second nearest (\jtwop) 
neighbour interactions
(\ie along [a,0,0] and [a,a,0], where a is the lattice constant), the parameters obtained for a lattice constant of 3.73~\AA~are
\jone = 10.7 meV, \jtwo = 0.3 meV and for a lattice constant of 3.75~\AA, the parameters are \jone = 10.1 meV, \jtwo = 0.3 meV. It is
generally believed that exchange interactions which connect magnetic ions along a linear chain are stronger than those which do not,
such as the \jtwo interaction here. However, in the cubic perovskite structure, exchange interactions along [2a,0,0], \etc contribute
equally to all four spin orderings studied and so cannot be extracted from the data presented here.
Similar values for \jone have been obtained
from model Hamiltonian calculations by Meskine \etal (\jone=~9.5 meV) \cite{Satpathy00}. Note that the definition used for
the exchange energy in that work, the difference between the energy of a pair of ferro- and anti-ferromagnetically coupled Mn ions,
is $\textit{twice}$ the exchange energy defined in Eq.~\ref{eqn:eqn1} above. Hence values for exchange energies
 from that work have been 
divided by two in order to compare them to values in the present work.

\subsection{\Cap: Cluster CI Calculations}

Exchange energies  obtained from cluster CI calculations depend strongly on Mn \eg and O 2p orbital populations.
In turn these populations depend on the Madelung potential of a sphere of point charges surrounding the \mcca cluster.
The charges were located on crystal ion sites and Mulliken populations of ions in bulk UHF calculations
on \Ca were used as a guide in choosing the magnitudes of these charges. The sphere of point charges had a radius
of just over 20~\AA~and contained around 3300 charges. The radius was chosen so that the sphere was overall almost charge 
neutral; each unit cell of point charges was also neutral. 
The dependence of exchange energies in \Ca and \La on Mn ion charge, 
measured by the Mulliken population of that ion in the SCF cluster calculation,
is shown in Fig. \ref{fig:fig5}. The charge on the two Mn ions $\textit{in the cluster}$ was adjusted by
transferring charge from Mn point charge sites to Ca or La point charge sites $\textit{in the sphere of point charges}$. The total
charge of the Mn and Ca (or La) point charges was kept constant and the O ion charge was maintained at the UHF Mulliken population value.
It can be seen that the magnitude of the exchange energy increases as charge is removed from the Mn ion, which is reasonable as
charge is mainly being transferred to/from the \eg orbitals which are directly involved in the exchange coupling mechanism - as the \eg
orbital becomes filled, the exchange energy diminishes. The CI calculation value of \jone=~8.1 meV 
quoted for \Ca in Table \ref{tab:tab2} is the value obtained
for an Mn cluster ion charge of +2.13, the Mn ion charge determined from the UHF calculation. 
This is to be compared to an estimate of the experimental value of \jone=~6.6 meV, derived from the N\'eel temperature of \Cap.

The fundamental SAF's for the septet and singlet states of the \mcca cluster were given in Eq. \ref{eqn:eqn2} and \ref{eqn:eqn4}. 
In the fundamental SAF wave functions for either spin state, each has a SAF coefficient, c$_{i}$, of unity,
however when additional
SAF's are permitted in the wave function (\ie permitting \oion ion 2p to \eg excitations, \etcp) the weights of fundamental SAF's are around
0.9950 and additional SAF's corresponding to O superexchange and \tg exchange enter the wave function with SAF coefficients of order 0.01.
Even for limited active spaces (as in these calculations) the number of SAF's entering the wavefunction means that a convenient way
to analyse the wavefunction is to tabulate the summed occupancies (\ie $\mid c_{i}^{2}\mid$) of configurations of a particular type. 
There are, for example, several 
SAF's in which one electron is excited from an O 2p orbital to an Mn \eg orbital \cite{cisqrd}. 
The relative magnitudes of these occupancies 
are a measure of the importance of each type of fluctuation about the fundamental SAF configurations.
Summed occupation numbers for the \mcca cluster are given in Table \ref{tab:tab3}. 
It can be seen that the fundamental (or main) SAF has occupancy 0.9926 for the singlet state while it has occupancy
0.9943 in the septet state and therefore that there are larger correlation effects in the singlet state. SAF's in which a \tg electron has
hopped from one Mn ion to the other have an occupancy of 0.0005, while these fluctuations are absent from the septet state 
owing to the Pauli exclusion principle,
as noted above. However, the main difference in septet and singlet wave functions is in the occupancy of states in which one electron is
transferred from an O 2p orbital to an \eg orbital, the occupancy being 0.0038 for the singlet state and 0.0027 for the septet state. 
The occupancy of SAF's in which a pair of electrons is transferred from O 2p to one Mn \eg orbital is the same for both spin states.
The energies of both spin states relative to the energy of the fundamental septet SAF are also given in Table \ref{tab:tab3}. The septet
state with O superexchange fluctuations is 133.4 meV below the fundamental septet SAF. This is the correlation energy for that state 
\cite{correlation}. The singlet state with O superexchange and \tg fluctuations is 149.6 meV below the reference energy and 153.2 meV
below the fundamental singlet SAF energy. The latter energy is the correlation energy for the singlet state.
Correlation energies for the \mcca and \mcla cluster CI wave functions are illustrated schematically in Figure \ref{fig:fig6}. 
Correlation energies are around 50 percent larger in \mcca than in \mcla and this is reflected in the larger exchange energy in \Cap.
It is worth noting that when the CI cluster calculation for the exchange energy in \Ca was performed with no point charge array
surrounding the cluster, the exchange energy obtained was 57 meV, well in excess of the experimental value. This emphasises the importance 
of Madelung terms in the crystal Hamiltonian in determining exchange energies in strongly correlated materials.

\subsection{\Lap: Bulk UHF Calculations}

Total energy calculations were performed on \La in the 
ideal perovskite (cubic) structure, a tetragonal perovskite structure, 
a cubic structure with a Jahn-Teller distortion of the MnO$_{6}$ octahedra 
and the \pnma structure with atomic coordinates derived from experiment \cite{Elemans71} and by minimising the total energy by varying
lattice parameters and internal coordinates not determined by symmetry. These structures are summarised in Table \ref{tab:tab4}. 
The Jahn-Teller distortion consisted of elongation or contraction
of Mn-O bonds parallel to the $ac$ axes of the unit cell.  These are the Mn-O bonds which induce FM coupling between Mn ions in the \pnma
structure. The cubic structure with the lowest energy had a lattice constant of 3.953 \AA~(volume 61.77~\AA$^{3}$ per Mn ion) which is 
comparable to the lattice constant of the 'cubic' phase of \La (3.947~\AA) which occurs at temperatures above 750K \cite{Moussa98}. 
All relative energies and lattice volumes will be assumed to be per Mn ion hereafter. When
this structure is changed by a 5 percent Jahn-Teller distortion (Table \ref{tab:tab4}) the energy is lowered by 304 meV 
and the magnetic ground state of the 
structure switches from \dxx\dzz FM to \dx\dy FM (see below). 

The total energy of the \pnma structure using coordinates from experiment \cite{Elemans71} (Table \ref{tab:tab4}) is 200~meV 
above the Jahn-Teller 
distorted structure. The total energy of the \pnma structure was minimised \cite{Lopt} by varying the lattice parameters and 7 internal 
coordinates not determined by symmetry of the \pnma space group. The total energy of the energy minimised structure is 6 meV above
the Jahn-Teller distorted structure. The optimised lattice parameters and internal coordinates are given in 
Table~\ref{tab:tab4}; the $a$ lattice vector is essentially unchanged while the $b$ and $c$ lattice vectors increase in magnitude 
by 1.1 and 1.6 percent, 
respectively. The lattice volume rises from 60.89 to 62.53~\AA$^{3}$.  Probably the most important changes which occur
on minimising the total energy are: the degree of Jahn-Teller distortion is reduced; La-O distances increase significantly. In the
lowest energy cubic structure there is one Mn-O distance of 1.976~\AA~and an La-O distance of 2.795~\AA.  On introducing the 5 
percent Jahn-Teller distortion these become Mn-O distances of 1.877, 1.976 and 2.075~\AA~and La-O distances of 2.795 and 2.797~\AA.
In the experimental \pnma structure \cite{Elemans71} the Mn-O distances are 1.903, 1.957 and 2.185~\AA~and 
the La-O distances are 2.433, 2.461 and 
2.548~\AA. These change to 1.910, 1.944 and 2.135~\AA~and 2.609, 2.666 and 2.684~\AA~in the energy minimised structure. Hence lower energies
are found for structures with larger La-O distances and a reduced Jahn-Teller distortion. The combined ionic radii of La$^{3+}$ and \oion
are 2.76 \AA \cite{Shannon76}.
La-O distances in the energy minimised \pnma and cubic structures lie just below the
combined ionic radii distance, whereas the La-O distances in the experimental \pnma structure lie well below this distance.
 The cubic structure with a
Jahn-Teller distortion and the energy minimised \pnma structure are both lower in energy than the lowest energy cubic structure by around 
300 meV. This energy lowering by a Jahn-Teller distortion is half of the lowering assumed by Millis
\cite{Millis96} in a calculation of electron-phonon coupling in \CaLap. 
The UHF calculations reported here are similar to those reported by Su \etal \cite{Su00}. They report an energy
lowering of 1055 meV when the cubic structure is changed to the experimental \pnma structure with no volume change. This calculation
will overestimate the energy difference between such structures as the cubic structure with the \pnma structure equilibrium volume
is not the mimimum energy cubic structure.

For the cubic perovskite structure it was found that variations of the total energies of different spin and orbital
orderings can be fitted very well by a Hamiltonian of the form

\begin{equation}\label{eqn:eqn7}
H = \sum_{<ij>}J_{ij} \frac{\hat{S}_{i}.\hat{S}_{j}}{S^{2}} + H_{OO}
\end{equation}

where $H_{OO}$ is an orbital ordering term which depends only on the orbital order.
For these calculations the cubic unit cell was doubled along [110], [101] and [011] directions (G-AF spin and orbital ordering)
and along the [001] direction (A-AF spin and orbital ordering)
and total energies and charge density difference plots \cite{cddiff} were
computed for \dxx\dxx, \dxx\dzz and \dzz\dzz orbital orderings and FM, A-AF and G-AF spin orderings. 
The \dxx\dzz A-AF combination
is incompatible with the unit cell doublings chosen and was omitted. 
Total energies are given in Table \ref{tab:tab5} and charge density difference plots for each of these
orbital orderings are shown in Fig.~\ref{fig:fig7}.

For \dxx\dxx and \dzz\dzz orbital ordering, distinct exchange constants in the $xy$ plane, \jparp, and in the $xz$ plane, \jperp,
are postulated, whereas for \dxx\dzz ordering a single exchange constant, J = \jpar = \jperp, is postulated.
Exchange constants for each orbital ordering 
are given in Table \ref{tab:tab6}. AF exchange constants are obtained when adjacent Mn orbital ordering is the same 
and FM coupling is observed when adjacent Mn \eg orbitals differ.
This observation also applies to \pnma structures studied: FM coupling is observed between in-plane Mn ions
with alternating \eg orbital orientations; AF coupling is observed when adjacent Mn \eg orbitals have the same orientation. 
Magnitudes of AF couplings
vary between 5.1 and 14.2 meV and one FM coupling of -6.0 meV is observed in the \dxx\dzz A-AF ordering. 

Once exchange constants have been computed for a particular orbital ordering, comparison of structures with the same magnetic structure
but different orbital ordering permits differences in orbital ordering energies to be calculated. The actual magnitude of orbital ordering
energy, $H_{OO}$, of course depends on the reference energy chosen. The choice of the FM~\dxx\dxx structure as the reference energy structure
yields values of -17.4, -20.0 and -113.4 meV for the \dxx\dxx, \dzz\dzz and \dxx\dzz relative orbital ordering energies. The important
result here is that an alternating orbital order (\dxx\dzz) is around 90 meV below those with the
same orbital order on each site (\dxx\dxx or \dzz\dzz) in the cubic perovskite structure. 
When the values of $H_{OO}$ and exchange constants just mentioned
are used to compute the relative energies of the eight spin and orbital orderings considered, the maximum deviation from the 
relative energies reported in Table \ref{tab:tab4} is 0.2 meV, demonstrating the suitability of the Hamiltonian in Eq. \ref{eqn:eqn7}.
The fact that charge density difference plots for different spin ordering and the same orbital ordering are very similar
suggests that this should be the case. 

Using the fact that orbital and spin contributions to the Hamiltonian are independent, differences in total energy of a
particular spin order as a function of lattice distortion may be attributed to differences in orbital ordering energy. Fig.~\ref{fig:fig8}
is a plot of total energy for each orbital ordering with G-AF magnetic order
as a function of isovolume, tetragonal lattice distortion. These calculations were performed using P4/mmm space group symmetry.
It can be seen that \dxx\dzz orbital ordering is the most stable ordering only within a small parameter range about the cubic structure.
When the tetragonal distortion is such that the lattice is elongated along the $z$ axis, 
\dzz\dzz ordering is favoured and when it is compressed along this axis, \dxx\dxx ordering is favoured. This may be explained by
a simple electrostatic argument - the ordering which is favoured in either case is the one where the occupied \eg orbitals are oriented
along the elongated axis or axes, thereby reducing the Coulombic repulsion energy. The greatest stabilisation relative to the cubic lattice
is found for an $x$/$z$ ratio of 0.94 where the energy is 164 meV below that of the cubic G-AF reference energy. This stabilisation
is still significantly less than the stabilisation of 298 meV which results when the energy minimised \pnma structure is adopted.

Relative energies and exchange constants for the Jahn-Teller distorted structure and both \pnma structures studied are given in Tables
\ref{tab:tab7} and \ref{tab:tab8}, respectively. Charge density difference plots for the Jahn-Teller distorted and energy minimised 
\pnma structures are shown in Fig.
\ref{fig:fig9}.
The magnetic ground state of the Jahn-Teller distorted structure is FM, but this is 
almost isoenergetic with the A-AF structure. This is because the in-plane exchange constant is FM while the out-of-plane exchange constant
is FM (but small). The magnetic ground state of the cubic structure is G-AF with AF coupling between all neighbouring Mn ions. The switch to
FM coupling between neighbouring Mn ions in-plane is due to the Jahn-Teller distortion in-plane. Both \pnma structures studied have A-AF 
magnetic ground states (as is the case in nature) but magnitudes of exchange constants obtained from these calculations are smaller
than those obtained from neutron scattering data \cite{Endoh96,Moussa96} (Table~\ref{tab:tab8}). Values of 0.6 and -3.7 meV for \jper and
\jpar may be compared to 0.8 and -3.5 meV obtained in a similar UHF calculation \cite{Su00} and 4.8 and -6.7 meV from experiment
\cite{Endoh96,Moussa96}. A local spin density approximation (LSDA) calculation \cite{Terakura96} found values of -3.1  and -9.1 meV for
\jper and \jpar. This calculation did find an A-AF ground state for \pnma \Lap, however, as second nearest neighbour exchange constants
are significant in the LSDA calculation and favour an A-AF magnetic ground state.

\subsection{\Lap: Cluster CI Calculations}

Cluster CI calculations for \La were performed using \mcla clusters with the Mn ions in the same configuration as a pair 
of Mn ions in the $ac$ plane (Fig. \ref{fig:fig1})
and with the Mn ions along a line parallel to the $b$ axis. The former cluster corresponds to a pair of Mn ions which is expected
to be FM coupled while the latter corresponds to a pair of ions which is expected to be AF coupled. 
Mn \eg orbital ordering in the former cluster had the form illustrated schematically in Fig. \ref{fig:fig2}c while the latter
had orbital ordering as in Fig \ref{fig:fig2}b.
Clusters and surrounding point charges with the experimental \pnma structure \cite{Elemans71} and the energy minimised structure 
were used.
Exchange constants for \La derived from these cluster calculations are given in Table~\ref{tab:tab8}. Cluster CI calculations
with Mn, O and La surrounding point charges of 2.6, -1.8 and 2.8 (close to Mulliken population values from UHF calculations) 
result in exchange constants
of 3.3 and -3.6 meV for \jper and \jpar when the experimental structure is used. These values change to 5.1 and -7.4 meV when the
energy minimised structure (Table \ref{tab:tab4}) is used.

The Madelung potential has an important role in determining exchange constants in manganites. 
Obviously ions several lattice constants
or more distant from the ions in the central cluster may be treated as point charges rather than distributed charges without 
significantly altering the potential within the central cluster. However point charges adjacent to the central cluster may cause
a significantly different potential within the cluster and affect the results of the exchange constant calculation. This question has
previously been addressed by other workers \cite{Martin93,Illas97}. In order to estimate the effect of terminating the cluster with
point charges, cluster CI calculations were performed with the 12 La point charges nearest to the central cluster ions 
replaced by La$^{3+}$ pseudopotentials \cite{Lapseudo}. This resulted in a
small increase in \jper and no change in \jpar compared to the calculation where only point charges were used. The values
obtained for \jper and \jpar from these calculations were 5.2 and -7.4 meV, which are in good agreement with the experimental values:
4.8 and -6.7 meV. Values for the exchange constants derived from the model Hamiltonian calculations of Meskine \etal \cite{Satpathy00}
are also given in Table~\ref{tab:tab8}.

Relative energies and SAF occupancies for the \mcla clusters used for calculating exchange constants in \La in the energy minimised structure
are given in Table~\ref{tab:tab9}. The fundamental SAF singlet states are 11.9 meV (\jperp) and 17.9 meV (\jparp) above the nonet states
of the clusters. When additional SAF's are permitted in the wave function the singlet(nonet) states are lowered by 105.4(83.3) meV 
(\jperp) and 82.8(79.7) meV (\jparp). These are the correlation energies for these states. The singlet state of the cluster used to 
calculate \jper is 10.2 meV lower in energy than the nonet state giving a value for \jper of 5.1 meV while the nonet state of the 
cluster used to calculate \jpar is 14.7 meV lower in energy than the singlet state giving a value of -7.4 meV for \jparp.
From Table~\ref{tab:tab9} it can be seen 
that O to \eg (1e) excitations are the main fluctuations about the fundamental SAF state. The weight of the fundamental SAF in the singlet 
states of either cluster is less than in the nonet states, reflecting the greater degree of correlation in the 
singlet states. In the \jper calculation the greater correlation energy of the singlet state \cf the nonet state is sufficient to make
the singlet state the ground state and give an AF exchange constant. On the other hand, in the \jpar calculation the singlet correlation 
energy is just greater than that of the nonet state and, together with the fact that the singlet state of the fundamental SAF wave function
is 17.9 meV above the nonet state, this results in a nonet ground state and an FM exchange constant.

\section{Discussion}
\label{sec:dis}

UHF and CI cluster calculations for the exchange constant in \Ca are in reasonable agreement with estimates for its value based on
the Rushbrooke-Wood formula \cite{Rushbrooke74} and its N\'eel temperature. 
The calculated exchange constants are larger than the estimate based 
on experiment. The single AF exchange constant is mainly a result of O to \eg (1e) excitations which lower the energy of the singlet state 
of a pair of adjacent Mn ions below that of the septet state. The magnitude of the exchange constant derived
from CI cluster calculations depends
strongly on the Madelung potential within the cluster and there is agreement between theory and estimates based on experiment only
when that potential results in ionic charges in the cluster similar to those in the bulk UHF calculation.

\La is more complex than \Cap. It is also more ionic than \Ca with Mulliken populations of ions closer to the formal ion charges. A number of 
orbital and spin ordered states exist within a small energy range, say 300 meV, close to the ground state. 
The energies of several spin and orbital ordered states of cubic \La are well
described by the Hamiltonian in Eq. \ref{eqn:eqn7}. In the remainder of this section exchange constants in cubic and Jahn-Teller distorted
\La are correlated with \mmm ion orbital ordering and \oion ion charge density distortions and \mmm ion interactions are identified
as type I, II or III according to Goodenough's scheme \cite{Goodenough55}. Finally, the role of correlation and availability
of empty orbitals on magnetic ion sites in AF and FM coupling is discussed.

Cubic \La has a \dxx\dzz FM
ground state and a lattice constant of 3.953 \AA. 
Exchange constants depend on orbital ordering and range from -6.0 to 14.2 meV. 
Charge density difference plots (Fig. \ref{fig:fig7}) show that the charge density on an Mn ion is essentially independent of charge densities
on neighbouring ions. That density is determined solely by the ion's orbital ordering. However, charge densities on more polarisable
\oion ion sites depend on charge densities at both neighbouring Mn ion sites. 
For \dzz\dzz orbital ordering (Fig. \ref{fig:fig7}, top panels), \oion ions in Mn-O bonds in the $xy$ plane undergo a 
quadrupolar distortion in which charge
is displaced from the Mn-O bond axis into directions perpendicular to the bond while \oion ions in Mn-O bonds along the $z$ axis are
much less severly distorted and the ions tend to elongate along the bond axes. The \dzz character of the ordered Mn \eg orbitals
can be seen clearly in the top right panel of Fig. \ref{fig:fig7}. The exchange constant for \mmm ions in the $xy$ plane with this
orbital ordering is \jpar = -0.1 meV while the exchange constant for \mmm ions along the $z$ axis is 14.2 meV. Thus a weak exchange
coupling is associated with the quadrupolar distortion of charge away from the bond axis while a much stronger coupling is associated
with a nearly spherical ion in which charge density tends to concentrate along the bond axis, compared to the spherical \oion ion.

For \dxx\dxx orbital ordering the \dxx character of the ordered Mn \eg orbitals is clearly seen in 
the middle left panel of Fig. \ref{fig:fig7}. There is a relatively weak quadrupolar distortion of the \oion ions in the $xy$ plane
and a stronger quadrupolar distortion of \oion ions along the $z$ axis. The exchange constant for \mmm ions in the $xy$ plane is 5.1 meV
while it is 7.2 meV for \mmm ions along the $z$ axis.

The \dxx or \dzz character of orbital ordering can be seen in both bottom panels in Fig. \ref{fig:fig7}. In the $xy$ and $xz$ planes
the \oion ion charge density is polarised in a circulating pattern, even though the \oion ions are situated
midway between the \mmm ions. Charge is polarised towards regions at \mmm ion sites where there is a reduction 
in charge density below that of spherical \mmmm ions, as indicated by negative contours in the charge density difference plots. 
Around each \mmm ion with the \dzz \eg orbital occupied, charge is deformed 
towards the \dzz ion in the $xy$ plane and away from it along the $z$ axis, whereas for \mmm ions with the \dxx \eg orbital occupied,
charge is deformed towards it along the $z$ axis and away from it in the $xy$ plane. Thus each \mmm ion is coupled to each neighbouring
\mmm ion by a polarised \oion ion and there is one FM exchange constant of -6.0 meV.

A simple pattern of orbital ordering is obtained for the Jahn-Teller distorted structure where the unit cell was doubled in 
the [001] direction (Fig. \ref{fig:fig9}). 
This pattern of orbital ordering was obtained without biasing the initial guess wave function in anyway (Appendix A).
Orbital ordering in the $xy$ plane is an alternating \dx \dy pattern which is repeated with period one along the $z$ axis.
This is the a-type orbital ordering discussed in Ref. \cite{Sawatzky99}.
The FM exchange constant between \mmm ions in the $xy$ plane is -8.1 meV and the weak FM exchange constant along the $z$ axis is -0.1 meV.
There is strong deformation of \oion ion charge density in the $xy$ plane towards regions of reduced 
charge density at the \mmm ion sites, associated
with strong FM exchange coupling (Fig. \ref{fig:fig9}, left panel). There is strong quadrupolar distortion
of the charge density at \oion ion sites coupling \mmm ions along the $z$ axis, associated with a weak FM exchange coupling. 

The same a-type orbital ordering is also found in the \pnma structures studied. 
There is a similar pattern of circulating charge polarisation towards
regions at \mmm ion sites where the charge density is reduced and there is strong FM exchange
coupling between \mmm ions lying approximately in the $ac$ plane (-3.7 and -6.0 meV, UHF calculations Table \ref{tab:tab8}). 

The three Mn-Mn interactions
described by Goodenough \cite{Goodenough55} are now tentatively identified in cubic and Jahn-Teller distorted \La using
charge densities on the \oion sites and 
orbital ordering at the neighbouring Mn ion sites. Type I interactions are found for: \dxx\dxx orbital ordering in cubic \La 
for both \jpar = 5.1 meV and
\jper = 7.2 meV 
(Fig. \ref{fig:fig7}, middle panels); 
\dzz\dzz orbital ordering for \jper = 14.2 meV
(Fig. \ref{fig:fig7}, top right panel).

Type II interactions are found for: \dxx\dzz orbital ordering in cubic \La  (Fig. \ref{fig:fig7}, bottom panels) for both \jper 
and \jpar = -6.0 meV; \dx\dy orbital ordering in Jahn-Teller distorted \La (Fig. \ref{fig:fig9}, left panel (\jpar =
-8.1 meV). Charge densities are characterised by breaking of symmetry of the \oion ion
along the Mn-O-Mn axis. 
Obviously this can only occur when the orbital orderings on adjacent Mn ions differ, however this observation is worth making
as such symmetry breaking is characteristic of FM exchange coupling. 

Type III interactions are found for \dzz\dzz orbital order in cubic \La (Fig. \ref{fig:fig7}, top left panel) where \jpar = -0.1 meV;
for \jper =-0.6 meV in the Jahn-Teller distorted structure (Fig. \ref{fig:fig9}, right panel).
In both cases the weak exchange coupling is associated with strong, quadrupolar \oion charge density deformation.

Cluster CI calculations provide detailed information on the exchange coupling mechanism. Fundamental SAF singlet states for clusters
representing both \Ca and \La lie above the fundamental SAF high spin multiplicity states; this is expected to be the case
for a wide range of magnetic ions exchange coupled \textit{via} a closed shell anion. 
The ground state for the pair of magnetic ions is AF when the \textit{difference} in correlation energies 
of the singlet and high spin multiplicity
states exceeds the singlet/high spin state splitting,
otherwise it is FM.
Correlation energies for singlet states
exceed correlation energies of the corresponding high spin multiplicity states in the three cases studied here (Fig. \ref{fig:fig6}).
This is also likely to be true for a wide range of magnetic ions which are exchange coupled \textit{via} a closed shell anion as there are
many more singlet SAF's than high spin SAF's in any particular active space. For example, in the active space used for the 
\La cluster CI calculations there are over 18,000 singlet SAF's compared to over 1,500 nonet SAF's, which simply reflects the fact that
there are many more ways to arrange spin-coupled electrons to form singlet states than there are to form nonet states for a specific
number of electrons. Only a few of either the singlet or nonet SAF's appear in the ground state wave functions with a signicant weight, 
but since there
are so many more singlet than nonet SAF's, it is not surprising that the singlet state correlation energy is larger. 

When one empty Mn \eg orbital is available to accept one or two electrons from an \oion ion, as is the case for \jpar in \pnma \Lap, 
the singlet state correlation energy is only slightly larger than the nonet state correlation energy (82.8 versus 79.7 meV) and the
nonet state is the ground state.  However, when two empty Mn \eg orbitals are available, as is the case for \jper in \pnma \La
and J in \Ca (Fig. \ref{fig:fig6}), singlet state correlation energies are significantly larger than the nonet(septet) state
correlation energies (105.4 versus 83.3 meV (\Lap) and 153.2 versus 133.4 meV (\Cap)) and the ground states are singlets.

Exchange coupling in \Ca and \La is largely due to quantum fluctuations in the ground state in which one electron is excited from an \oion ion
into an Mn \eg orbital. Fluctuations in which an electron is exchanged between \tg orbitals enter the singlet state in \mcca and \mcla
clusters but are not the main contributors to the exchange interaction. Parallel studies of exchange coupling in \LaCu
\cite{Zheng01} show that \dxx\dxx exchange interactions dominate the exchange coupling in \LaCu~and O to \dxx excitations have
a lesser weight than Cu$^{1+}$Cu$^{3+}$ excitations in the \LaCu~ground state. 
This difference in exchange coupling mechanism most likely reflects trends in effective
Hubbard U parameters for Mn$^{3+/4+}$ and Cu$^{2+}$ and O 2p to metal d excitation energies.

\acknowledgments
This work was supported by Enterprise Ireland under grant number SC/00/267.  M.~N. wishes to acknowledge support by the Trinity Trust.

\appendix

\section{Details of Calculations}
\label{app:details}

The methods used to generate localised orbitals and point charge arrays and the basis sets and CI computer codes
used in this work are described in this section. 

UHF calculations on the crystalline solid were performed using the \textsc{crystal98} code \cite{crystal98}. The basis sets used for both
crystal UHF calculations and cluster CI calculations were identical Gaussian orbital basis sets designed for use in the solid state.
They are slightly modified versions of the basis sets available from the \textsc{crystal98} website \cite{crystalwebsite}. Outer 
exponents of the Gaussian functions were modified so that the total energy in a UHF calculation on \Ca was minimised.
The original basis sets had been optimised for different Mn ionic solids.
The basis sets used in all calculations are the all electron basis sets 
for: Mn (86-411d41G \cite{Towler94} with two d orbital exponents, optimised for \Ca by changing the outer d exponent to 0.259);
O (8-411G \cite{Towler94} with principal quantum number up to n = 4, optimised for \Ca by changing the outer sp exponents
to 0.4763 and 0.22);
Ca (86-511d3G \cite{Mackrodt93} with the outer d orbital exponents optimised for \Ca to 3.191, 0.8683 and 0.3191) and
an La basis set optimised for the La$^{3+}$ ion \cite{Towler96}. The La basis set used in this work differs from the cited basis in that
the 5d orbital was removed from the basis and the 6sp and 7sp orbitals were replaced by a single sp orbital exponent of 0.3917.

Different orbital ordered states in UHF calculations were obtained using a feature in the \textsc{crystal98} code which increases
the diagonal element of the Fock matrix
corresponding to a particular orbital for the first few iterations of the calculation. This results in that orbital being unoccupied
during those SCF cycles and allows
the wave function to converge to a state which is a local energy minimum with a particular orbital ordering.

High spin multiplicity states, such as the septet and nonet states of the clusters used here, are generally well described by a self
consistent field (SCF) restricted open shell Hartree-Fock (ROHF) wave function. 
All electrons on the cluster were treated explicitly - no pseudopotential approximation was used,
except in the test calculation with a La$^{3+}$
pseudopotential described above.
CI calculations were performed in localised orbital bases, rather than the 
canonical molecular orbital bases derived from the SCF ROHF calculations.
Localisation of SCF ROHF molecular orbitals was performed using the Foster-Boys algorithm \cite{Boys62}, which generates
localised orbitals with maximally separated centroids. Doubly occupied O 2p orbitals,
singly occupied Mn \tg (or \tg and an \eg orbital for \mclap) orbitals and unoccupied Mn \eg orbitals were localised in three 
separate localisation steps. These must be performed separately in order to preserve invariance of the ROHF total energy, since any
mixing between orbitals of different occupancy will increase the total energy. 
In the localised orbital ROHF wave functions for either spin state of the \mcca and \mcla clusters, 
each Mn d electron occupies a separate orbital.

Calculations on low spin multiplicity states 
of the clusters used the same sets of localised orbitals. 
They demonstrate that the localised orbitals generated
for the high spin multiplicity states are very good approximations to the optimal orbitals for open shell low spin multiplicity states
and that a high spin multiplicity ROHF wave function ought to be an excellent starting point for perturbative calculations
on high and  low spin multiplicity states in the solid state. 

In a CI calculation on a cluster of this size it is essential to partition the 
orbital space into a core space (with doubly occupied orbitals), an active space of orbitals which are occupied or unoccupied in the
ROHF main configuration and a space of redundant, unoccupied orbitals which are not used in the calculation. The active orbitals in 
this work were the three O 2p localised orbitals on the central \oion and a set of \tg and \eg orbitals on each Mn ion. The 
Direct Multi-Reference CI module \cite{MRDCI} in the GAMESS \cite{GAMESS} programme was
used for this work. The active space consisted of either the (single) fundamental SAF orbitals or
the fundamental SAF plus all possible single or double excitations which can be made from the fundamental SAF into empty active
space orbitals. 

Calculations were performed for clusters with no surrounding point charges and with point charges in a spherical volume surrounding the
cluster. The radius of the sphere was over 20 \AA~and included around 3300 charges. The charges were located on the ionic
sites of either \Ca or \Lap. Mulliken populations derived from UHF crystal calculations were used as guides for
point charge magnitudes.
For \Cap, UHF Mulliken populations were Ca$^{+1.86}$Mn$^{+2.13}$O$^{-1.33}_{3}$.
However, in the SCF ROHF cluster calculation, this choice of point charges results in 
Mulliken populations of Mn$^{+2.60}$O$^{-1.31}$Mn$^{+2.60}$
on the central Mn-O-Mn chain in the cluster. The Mn and Ca point charge magnitudes were adjusted to Ca$^{1.15}$Mn$^{2.84}$O$^{-1.33}$
and this resulted in Mulliken populations of Mn$^{+2.17}$O$^{-1.61}$Mn$^{+2.17}$ on the central Mn-O-Mn chain and populations of -1.64 and
-1.67 on the other two O types in the cluster. Note that this adjustment leaves each point charge unit cell charge neutral 
and the point charge
sphere radius is adjusted so that the entire cluster has a charge near zero. 
The major changes which occur on adjusting the point charges are: charge is transferred from the outer O ions in the cluster to the 
Mn ions and central O ion, each gaining about 0.4e; the AF exchange constant changes from 21.0 meV to 8.1 meV, in agreement 
with other calculation methods and in reasonable agreement with experiment; 
the degree of correlation in the wavefunction decreases sharply.
When a \mcca cluster with no external point charges is used,
the Mulliken populations on the central Mn-O-Mn chain are
Mn$^{+2.46}$O$^{-0.94}$Mn$^{+2.46}$ and the exchange coupling energy is 57 meV.

A similar adjustment of point charge magnitudes was used for the \mcla cluster CI calculations.
The Mulliken populations determined from UHF crystal calculations on the experimental \pnma structure were
La$^{+3.15}$O$^{-1.75,-1.82}$Mn$^{+2.24}$. Cluster point charges of  
La$^{+2.80}$O$^{-1.80,-1.80}$Mn$^{+2.60}$ resulted in Mulliken populations of
Mn$^{+2.45}$O$^{-1.65}$ in the \pnma structure.

%%%%%%%%%%%%%%%%%%%%%%%%%%%%%%%%%%%%%

%%%%%%%%%%%%%%%%%%%%%%%

%\begin{flushleft}

\begin{figure}[!ht]
\caption{\pnma structure of \La according to Elemans \cite{Elemans71}. 
Mn-O bonds are shown explicitly. Mn ions are dark spheres, O ions are light spheres and La ions are 
unconnected light spheres. Mn ions labelled 1 and 2 are AF coupled (\jperp) and Mn ions labelled 2 and 3 are FM coupled (\jparp).
The cluster used to compute the AF coupling constant had the same structure as Mn ions 1 and 2 and their associated \oion ion quasi-octahedra.
The cluster used to compute the FM coupling constant had the same structure as Mn ions 2 and 3 and their associated \oion ion quasi-octahedra.}
\label{fig:fig1}
\end{figure}

\begin{figure}[!ht]
\caption{Empty orbital ordering in \Lap. (A) The empty orbital arrangement which results when occupied orbitals are \dxx~\dxx ordered.
(B) Empty orbital arrangement with AF spin coupling favoured, (C) empty orbital arrangement with FM spin coupling favoured, (D) empty
orbital arrangement with weak spin coupling.}
\label{fig:fig2}
\end{figure}

\begin{figure}[!ht]
\caption{Localised orbital basis used for \Ca cluster CI calculations. 
Top panel: O 2p$_z$ orbital; middle panel: Mn d$_{xz}$ orbital; bottom panel:Mn \dzz orbital. 
The latter is the empty \eg orbital responsible for exchange coupling.}
\label{fig:fig3}
\end{figure}

\begin{figure}[!ht]
\caption{Localised orbital basis used for \jper exchange constant cluster CI calculation for \Lap. 
Top panel O 2p$_{z}$ orbital; middle panel: filled \eg orbital perpendicular to Mn-O-Mn axis; 
bottom panel: empty \eg orbital oriented along Mn-O-Mn axis.}
\label{fig:fig4}
\end{figure}

\begin{figure}[!ht]
\caption{Exchange coupling constants for \Ca and \La from CI cluster calculations with varying Mn ion Mulliken populations. The variation
in Mulliken population was induced by changing the magnitude of point charges at Mn and La or Ca ion sites.}
\label{fig:fig5}
\end{figure}

\begin{figure}[!ht]
\caption{Correlation energies in \Ca and \Lap. Magnitudes of correlation energies are illustrated by vertical arrows and are given in meV
in plain text. Energy differences between low and high multiplicity spin states are given in italics and occupancies of the 
fundamental SAF in each state are given at the base of each arrow. The horizontal line is the SCF ROHF energy for each state.}
\label{fig:fig6}
\end{figure}

\begin{figure}[!ht]
\caption{Charge density difference plots for cubic \La with (a,b) \dzz\dzz, (c,d) \dxx\dxx, (e,f) \dxx\dzz orbital ordering. 
Left and right panels show density differences in the $xy$ and $xz$ planes, respectively.
The differences in charge densities are the UHF SCF density for the solid minus the UHF SCF densites for the \oion ions and the \mmmm ion.}
\label{fig:fig7}
\end{figure}

\begin{figure}[!ht]
\caption{Total energies of \dxx\dxx, \dzz\dzz and \dxx\dzz orbital ordered structures with G-AF magnetic ordering. The reference energy
is the cubic \La \dxx\dxx FM energy (Table \ref{tab:tab5}).}
\label{fig:fig8}
\end{figure}

\begin{figure}[!ht]
\caption{Charge density difference plots for \La with a 5 percent Jahn-Teller distortion in the $xy$ plane.
Panels (a) and (b) show density differences in the $xy$ and $xz$ planes, respectively.
The differences in charge densities are the UHF SCF density for the solid minus the UHF SCF densites for the \oion ions and the \mmmm ion.}
\label{fig:fig9}
\end{figure}

%%%%%%%%%%%%%%%%%%%%%%%

\begin{table}[!tbp]
\caption{Relative energy and magnetic moment per Mn ion in \Cap.}
\begin{tabular}{c d d d d d}
			Spin Ordering
		&	Relative Energy (meV)\tablenotemark[1]
		&	$\mu$($\mu_{B})$\\
\hline
FM				&	0.0	&	 3.00\\
A-AF				&     -23.7	&	 3.27\\
C-AF				&     -45.1	&	 3.15\\
G-AF				&     -64.3	&	 3.23\\
\end{tabular}\label{tab:tab1}
\tablenotemark[1]{Lattice constant 3.73 \AA}
\end{table}

%%%%%%%%%%%%%%%%%%%%%%%

\begin{table}[!tbp]
\caption{Exchange constants in \Ca derived from experiment and \ab calculations.}
\begin{tabular}{c d} 
					&      	J(meV)\\
\hline
Experiment\tablenotemark[1]		&	 6.6\\
Cluster CI\tablenotemark[2]		&	 8.1\\
Model Hamiltonian\tablenotemark[3]	&	 9.5\\
UHF\tablenotemark[2]     		&	10.7\\
\end{tabular}
\label{tab:tab2}
\tablenotemark[1]{Rushbrooke \etal \cite{Rushbrooke74}}\\
\tablenotemark[2]{This work. Lattice constant 3.73 \AA}\\
\tablenotemark[3]{Meskine \etal \cite{Satpathy00}}\\
\end{table}

%%%%%%%%%%%%%%%%%%%%%%%

\begin{table}[!tbp]
\caption{Relative energy and SAF occupation numbers for singlet and septet states of \mcca cluster representing \Cap.}
\begin{tabular}{c d d d d d}
			State
		&	Energy (meV)\tablenotemark[1]	
		&	Main SAF
		&	\tg~ Exchange
		&	O to \eg~(1e) 
		&	O to \eg~(2e)\\ 
\hline
singlet\tablenotemark[2]	&	+3.6	&	1.0000	&	0.0000		&	0.0000		&	0.0000\\
septet\tablenotemark[2]		&	0	&	1.0000	&	0.0000		&	0.0000		&	0.0000\\
singlet\tablenotemark[3]	&	-149.6	&	0.9926	&	0.0005		&	0.0038		&	0.0017\\
septet\tablenotemark[3]	        &	-133.4	&	0.9943	&	0.0000		&	0.0027		&	0.0017\\
\end{tabular}\label{tab:tab3}
\tablenotemark[1]{Energies are relative to the restricted open shell Hartree-Fock septet state}\\
\tablenotemark[2]{Fundamental SAF only}\\
\tablenotemark[3]{Fundamental SAF + all single and double excitations in active space from fundamental SAF}\\
\end{table}

%%%%%%%%%%%%%%%%%%%%%%%

\begin{table}[!tbp]
\caption{Structural parameters in Jahn-Teller distorted \La and \pnma \La determined by experiment and total energy minimisation.
Each cell is a $\sqrt{2}$x2x$\sqrt{2}$ doubling of the primitive perovskite unit cell.}
\begin{tabular}{c d d d}
Ion		&	x	&	y	&	z\\
\hline
La\tablenotemark[1]	&       0.549	&	0.250	&	0.010\\
Mn			&	0.000	&	0.000	&	0.000\\
O			&	-0.014	&	0.250	&	-0.070\\
O			&	0.309	&	0.039	&	0.224\\
\hline
La\tablenotemark[2]	&       0.517	&	0.250	&	0.001\\
Mn			&	0.000	&	0.000	&	0.000\\
O			&	-0.002	&	0.250	&	-0.027\\
O			&	0.290	&	0.014	&	0.237\\
\hline
La\tablenotemark[3]	&       0.500	&	0.250	&	0.000\\
Mn			&	0.000	&	0.000	&	0.000\\
O			&	0.000	&	0.250	&	0.000\\
O			&	0.2625	&	0.000	&	0.2625\\
\end{tabular}\label{tab:tab4}
\tablenotemark[1]{Experimental structure \cite{Elemans71} (Fig. \ref{fig:fig1}), lattice parameters a=5.742\AA, b=7.668\AA, c=5.532\AA}\\
\tablenotemark[2]{Optimised structure, lattice parameters a=5.740\AA, b=7.754\AA, c=5.620\AA}\\
\tablenotemark[3]{Jahn-Teller distorted structure, lattice parameters a=5.590\AA, b=7.905\AA, c=5.590\AA. Note that the Jahn-Teller
distortion is in the $xz$ plane in this Table to allow easy comparison between its structural parameters and those of the \pnma structures.
Elsewhere in this work the Jahn-Teller distortion is assumed to be in the $xy$ plane.}\\
\end{table}

%%%%%%%%%%%%%%%%%%%%%%%

\begin{table}[!tbp]
\caption{Relative energy and magnetic moment per Mn ion in cubic \La with various spin and orbital orderings.}
\begin{tabular}{c d d d d d}
			Spin and Orbital Ordering\tablenotemark[1]
		&	Relative Energy (meV)
		&	$\mu$($\mu_{B})$\\
		&
		&\\
\hline
FM \dxx\dxx			&       0.0	&	 4.00\\
FM \dxx\dzz			&    -131.5	&	 4.00\\
FM \dzz\dzz			&      -6.1     &	 3.99\\
A-AF \dxx\dxx			&     -14.4	&	 4.05\\
%A-AF \dxx\dzz			&     -40.2	&	 3.87\\
A-AF \dzz\dzz			&     -34.4	&	 3.87\\
G-AF \dxx\dxx			&     -34.9	&	 3.88\\
G-AF \dxx\dzz			&     -95.4	&	 3.89\\
G-AF \dzz\dzz			&     -34.0	&	 3.88\\
\end{tabular}\label{tab:tab5}
\tablenotemark[1]{Lattice constant 3.934\AA}\\
\end{table}

%%%%%%%%%%%%%%%%%%%%%%%

\begin{table}[!tbp]
\caption{Exchange constants in cubic \La with various orbital orderings.}
\begin{tabular}{c d d} 
					Spin and Orbital Ordering\tablenotemark[1]
					&	\jper(meV)\tablenotemark[2]&	\jpar(meV)\tablenotemark[3]\\
\hline
\dxx\dxx				&	7.2	&	5.1\\
\dxx\dzz				&	-6.0	&	-6.0\\
\dzz\dzz				&	14.2	&	-0.1\\
\end{tabular}
\label{tab:tab6}
\tablenotemark[1]{Lattice constant 3.953\AA}\\
\tablenotemark[2]{Exchange constant for Mn ions coupled perpendicular to $ac$ plane}\\
\tablenotemark[3]{Exchange constant for Mn ions coupled parallel to $ac$ plane}\\
\end{table}

%%%%%%%%%%%%%%%%%%%%%%%%%%%%%%%%%%%

\begin{table}[!tbp]
\caption{Relative energy and magnetic moment per Mn ion in \pnma and Jahn-Teller distorted cubic \Lap.}
\begin{tabular}{c d d d d d}
                        Structure and Spin Ordering	& Energy(meV)	&	$\mu$($\mu_{B}$)\\
\hline
\pnma (Experiment) FM\tablenotemark[1]				&       0.0     &	 4.00\\
\pnma (Experiment) A-AF						&      -1.2     &	 4.00\\
\pnma (Experiment) G-AF						&      13.9     &	 3.96\\
\hline
\pnma (Optimised) FM\tablenotemark[2]				&       0.0     &	 4.00\\
\pnma (Optimised) A-AF						&      -2.0     &	 3.96\\
\pnma (Optimised) G-AF						&      21.9     &	 3.94\\
\hline
Jahn-Teller FM\tablenotemark[3]					&       0.0     &	 4.00\\
Jahn-Teller A-AF						&       1.1     &	 3.98\\
Jahn-Teller G-AF						&      33.6     &	  - \\
\end{tabular}\label{tab:tab7}
\tablenotemark[1]{Reference energy is 194 meV above optimised FM \pnma structure (Table~\ref{tab:tab4}).}\\
\tablenotemark[2]{Reference energy is that of this structure and magnetic order (Table~\ref{tab:tab4}).}\\
\tablenotemark[3]{Reference energy is 8 meV below optimised FM \pnma structure (Table~\ref{tab:tab4}).}\\
\end{table}

%%%%%%%%%%%%%%%%%%%%%%%

\begin{table}[!tbp]
\caption{Exchange constants in \pnma \La derived from experiment and \ab and model Hamiltonian calculations.}
\begin{tabular}{c d d} 
					&	\jper(meV)\tablenotemark[1]&	\jpar(meV)\tablenotemark[2]\\
\hline
Experiment\tablenotemark[3]				&	 4.8	&	-6.7\\
UHF(Experiment)\tablenotemark[4]			&	 0.6	&	-3.7\\
UHF(Experiment)\tablenotemark[5]     			&	 0.8	&	-3.5\\
UHF(Optimised)\tablenotemark[6]				&	 1.0	&	-6.0\\
UHF(Jahn-Teller)\tablenotemark[7]			&	 -0.6	&	-8.1\\
LSDA(Experiment)\tablenotemark[8]  			&	-3.1	&	-9.1\\
Cluster CI(Experiment)\tablenotemark[4]			&	 3.3	&	-3.6\\
Cluster CI(Optimised)\tablenotemark[6]			&	 5.1	&	-7.4\\
Cluster CI(Optimised/La pseudopotential)\tablenotemark[6]&	 5.2	&	-7.4\\
model Hamiltonian\tablenotemark[9]			&	 4.8	&	-3.7\\
\end{tabular}
\label{tab:tab8}
\tablenotemark[1]{Exchange constant for Mn ions coupled perpendicular to $ac$ plane}\\
\tablenotemark[2]{Exchange constant for Mn ions coupled parallel to $ac$ plane}\\
\tablenotemark[3]{Hirota \etal \cite{Endoh96};Moussa \etal \cite{Moussa96}}\\
\tablenotemark[4]{This work. Elemans structure \cite{Elemans71}(Table~\ref{tab:tab4}).}\\
\tablenotemark[5]{Su \etal \cite{Su00}.}\\
\tablenotemark[6]{This work. Optimised structure (Table~\ref{tab:tab4}).}\\
\tablenotemark[7]{This work. Jahn-Teller distorted structure (Table~\ref{tab:tab4}).}\\
\tablenotemark[8]{Soloyev \etal \cite{Terakura96}.}\\
\tablenotemark[9]{Meskine \etal \cite{Satpathy00}}\\
\end{table}

%%%%%%%%%%%%%%%%%%%%%%%%%%%%%%%%%%%

\begin{table}[!tbp]
\caption{Relative energy and orbital occupation numbers for singlet and nonet states of \mcla cluster representing \Lap.}
\begin{tabular}{c d d d d d}
			State
		&	Energy (meV)\tablenotemark[1]	
		&	Main SAF
		&	\tg~ Exchange
		&	O to \eg~(1e) 
		&	O to \eg~(2e)\\ 
\hline
singlet\tablenotemark[2]\tablenotemark[4]&	+11.9	&	1.0000	&	0.0000		&	0.0000		&	0.0000\\
nonet\tablenotemark[2]\tablenotemark[4]&	  0.0	&	1.0000	&	0.0000		&	0.0000		&	0.0000\\
singlet\tablenotemark[3]\tablenotemark[4]&	-93.5	&	0.9937	&	0.0006		&	0.0037		&	0.0007\\
nonet\tablenotemark[3]\tablenotemark[4]&	-83.3   &	0.9954	&	0.0000		&	0.0030		&	0.0007\\
\hline
singlet\tablenotemark[2]\tablenotemark[5]&	+17.9	&	1.0000	&	0.0000		&	0.0000		&	0.0000\\
nonet\tablenotemark[2]\tablenotemark[5]&	  0.0	&	1.0000	&	0.0000		&	0.0000		&	0.0000\\
singlet\tablenotemark[3]\tablenotemark[5]&	-64.9	&	0.9949	&	0.0004		&	0.0025		&	0.0006\\
nonet\tablenotemark[3]\tablenotemark[5]&	-79.9   &	0.9946	&	0.0000		&	0.0038		&	0.0008\\
\end{tabular}\label{tab:tab9}
\tablenotemark[1]{Energies are relative to the restricted open shell Hartree-Fock septet for the whole cluster, \ie per Mn ion pair.}\\
\tablenotemark[2]{Fundamental SAF only}\\
\tablenotemark[3]{Fundamental SAF + all single and double excitations in active space from fundamental SAF}\\
\tablenotemark[4]{\jper calculation}\\
\tablenotemark[5]{\jpar calculation}\\
\end{table}

%%%%%%%%%%%%%%%%%%%%%%%%%%%%%%%%%%%%%%%%%%%

\begin{thebibliography}{42}
\bibitem{fermion}
Fermion eigenstates of a Hamiltonian which does not explicitly contain spin may be written as anti-symmetric products of spatial orbitals
and spin eigenfunctions (SEF's)\cite{Pauncz}. SEF's are eigenfunctions of the total spin operator, $\hat{S}^{2}$, and $\hat{S}_{z}$ operator.
SEF's may be generated in several different ways and the Multi-Reference Doubles CI module \cite{MRDCI} of the 
GAMESS \cite{GAMESS} code used in this work uses the Yamanouchi-Kotani (YK)
scheme \cite{Yamanouchi36}. There is one SEF for six electrons coupled into a septet state while there are five SEF's for
six electrons coupled into a singlet state. Each SEF consists of an orthonormal combination of products of the 
one-electron eigenspinors, $\alpha$ or $\beta$.
\bibitem{cisqrd}
Sums of SAF occupancies are slightly less than unity for states other than the fundamental SAF state because large numbers of
SAF's with coefficients smaller than 0.003, which individually contribute 0.000009 to the total SAF occupancy, were omitted from the sums.
\bibitem{correlation}
The correlation energy for a molecule, cluster, \etc is usually defined to be the energy difference between that of an SCF wave function
and a correlated wave function.  Here the correlation energy is defined to be the difference between the fundamental
SAF energy for each state and the energy when fluctuations within the limited active space described above are taken into account. Additional
correlation energies, for example those due to fluctuations within an O ion, will be larger.
However, it is believed that the correlation effects included in this
work are the most important in determining exchange constants.
\bibitem{cddiff}
These are plots of the difference in the charge density obtained from the UHF wave functions of the solid and
superimposed (spherical) densities of \mmmm ions with half-filled \tg shells and \oion ions. The plots therefore show the charge density
of the Mn \eg electron and the distortion in density of an \oion ion in the solid state.
\end{thebibliography}
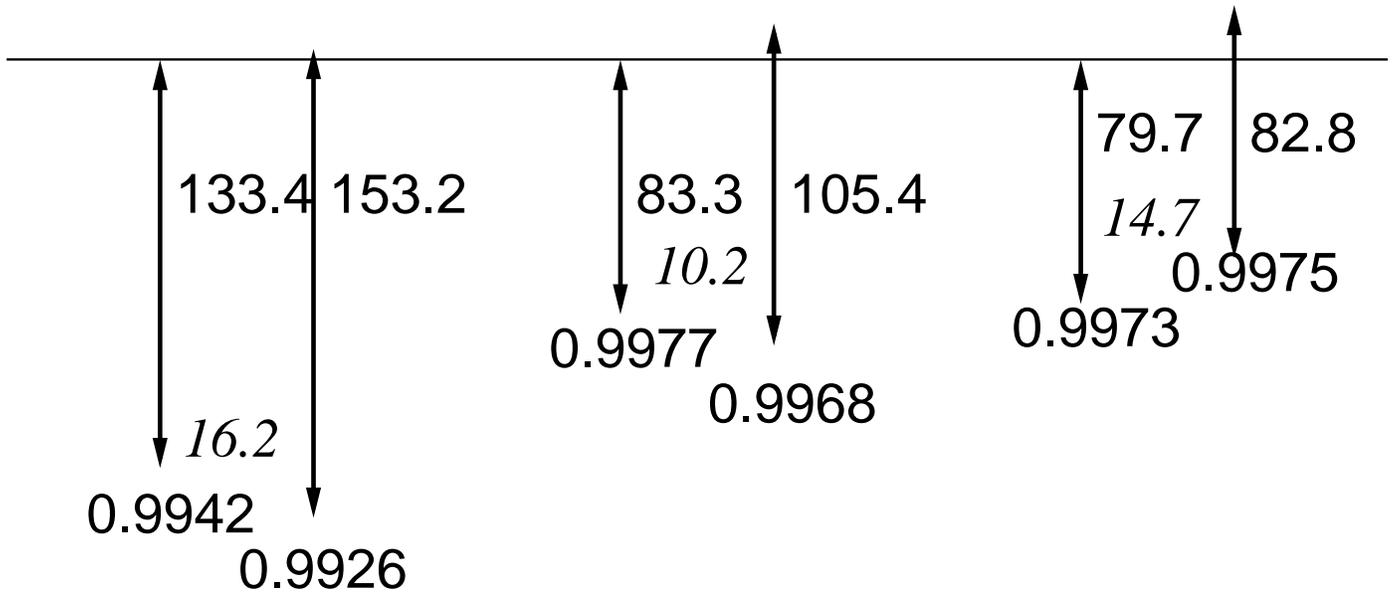
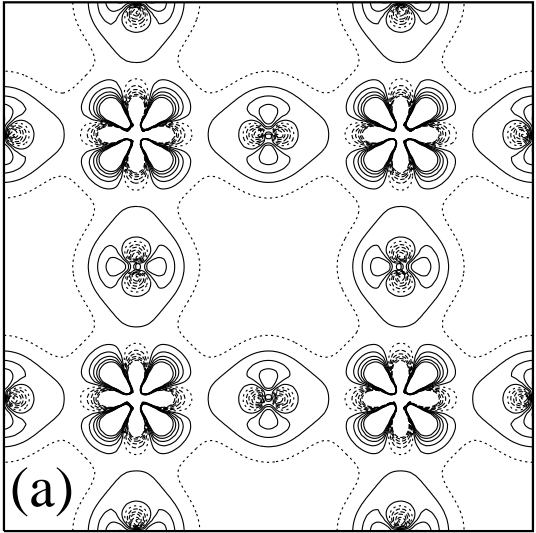
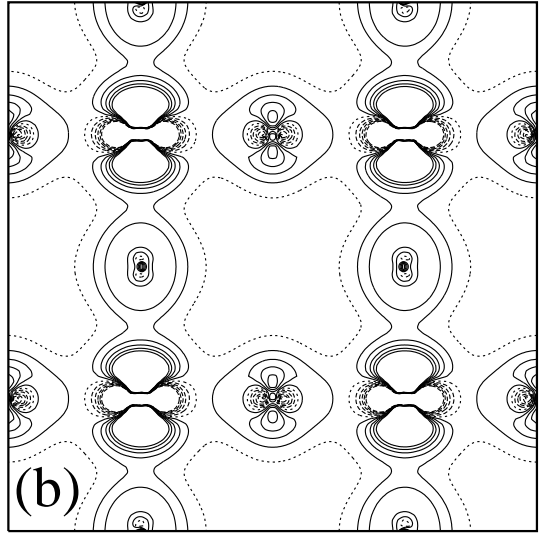
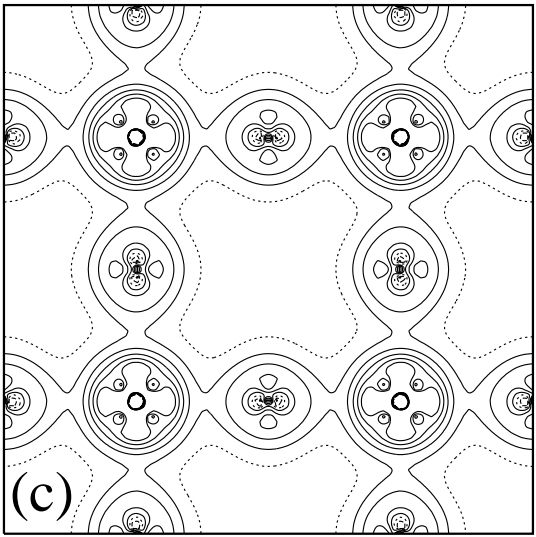
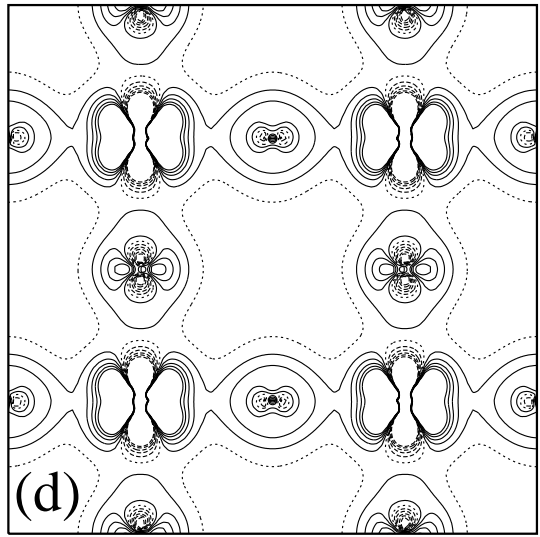
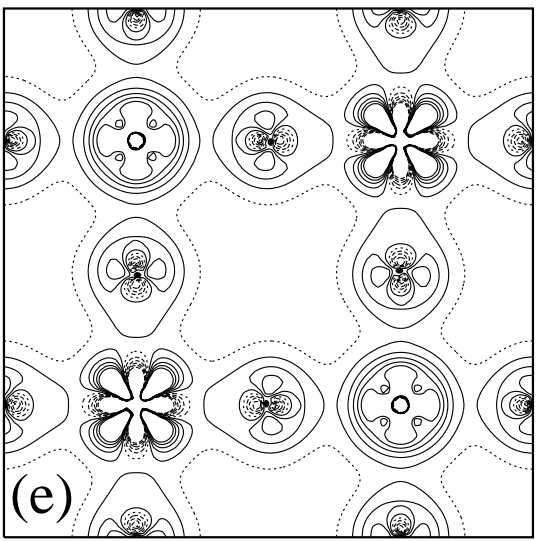
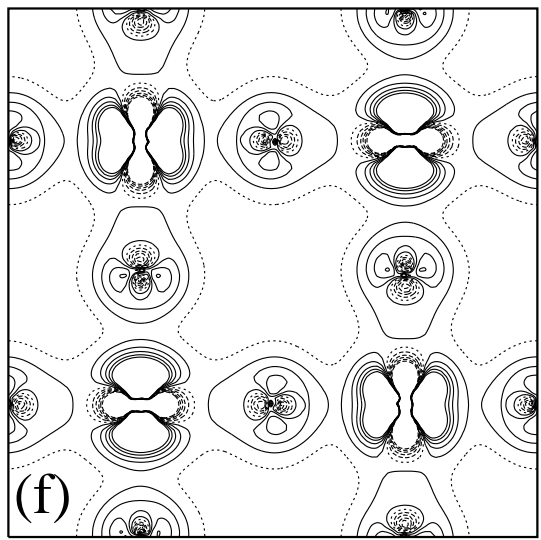
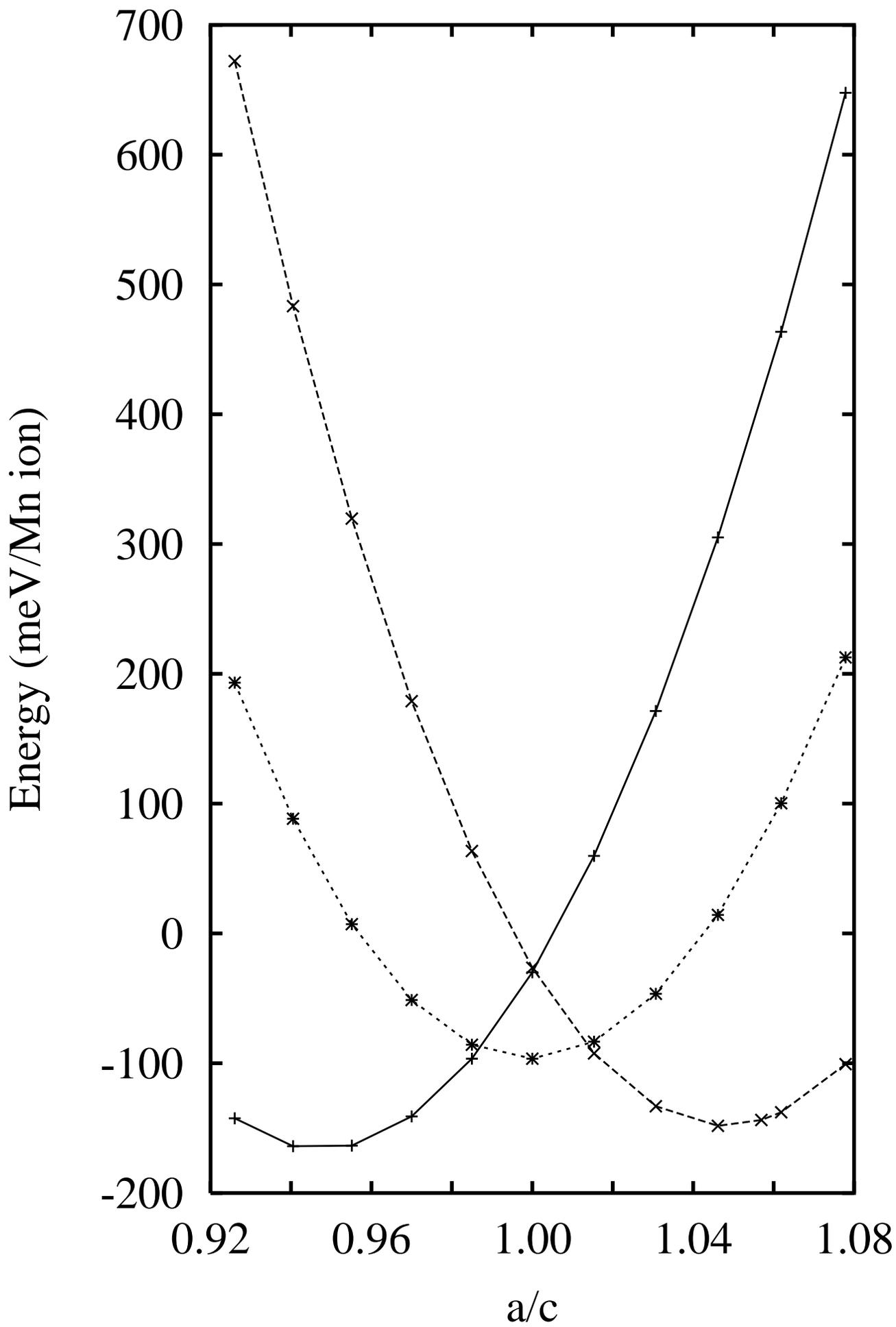
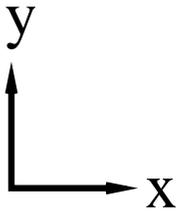
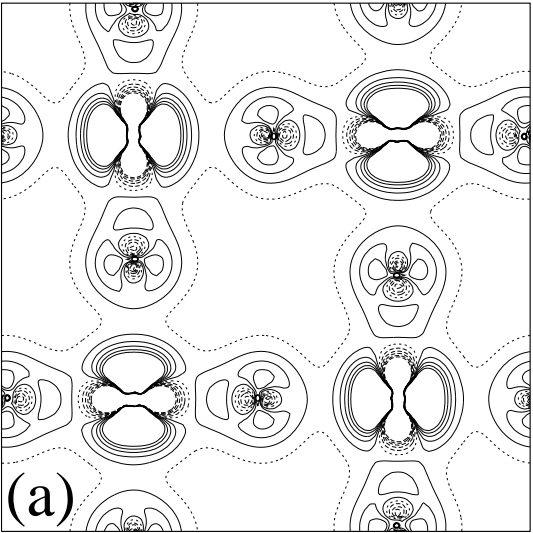
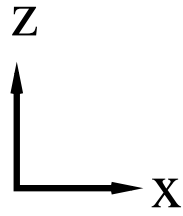
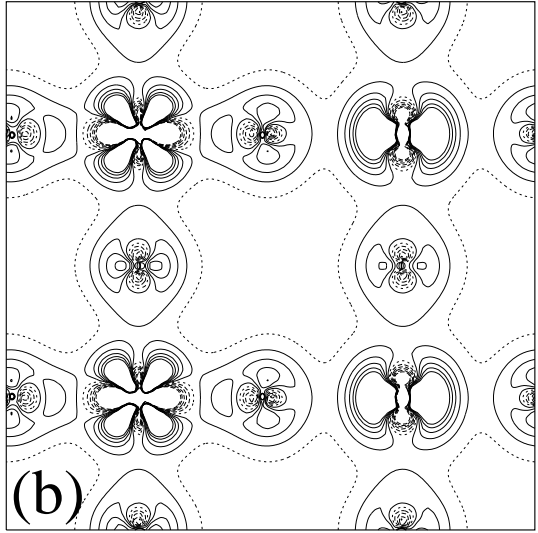
\end{document}